\documentclass[conference, a4paper, 10pt]{IEEEtran}
\IEEEoverridecommandlockouts
\usepackage{amsmath,amssymb,amsfonts}
\usepackage{subcaption}
\usepackage{bm, bbm}
\usepackage{graphicx}
\usepackage{textcomp}
\usepackage{xcolor}
\usepackage{url}
\usepackage{hyperref}
\usepackage{algorithm}
\usepackage{algpseudocode}
\usepackage[backend=biber,
sorting=none,
doi=false,
isbn=false,
url=false,
maxnames=6,
minnames=1,
style=ieee,
citestyle=numeric-comp]{biblatex}
\newcommand{\cnew}{c_{\text{new}}}
\newcommand{\cold}{c_{\text{old}}}
\DeclareMathOperator{\sinc}{sinc}
\begin{document}

\title{Online Single-Channel Audio-Based Sound Speed Estimation for Robust Multi-Channel Audio Control}

\author{\IEEEauthorblockN{Andreas~Jonas~Fuglsig, Mads~Græsbøll~Christensen and Jesper~Rindom~Jensen} \IEEEauthorblockA{\textit{Department of Electronic Systems,}\\ \textit{Aalborg University, 9220 Aalborg, Denmark}\\ Email: \{ajf,mgc,jrj\}@es.aau.dk}}

\maketitle

\begin{abstract} 
Robust spatial audio control relies on accurate acoustic propagation models, yet environmental variations, especially changes in the speed of sound, cause systematic mismatches that degrade performance. Existing methods either assume known sound speed, require multiple microphones, or rely on separate calibration, making them impractical for systems with minimal sensing. We propose an online sound speed estimator that operates during general multichannel audio playback and requires only a single observation microphone. The method exploits the structured effect of sound speed on the reproduced signal and estimates it by minimizing the mismatch between the measured audio and a parametric acoustic model. Simulations show accurate tracking of sound speed for diverse input signals and improved spatial control performance when the estimates are used to compensate propagation errors in a sound zone control framework. 
\end{abstract}


Spatial audio control techniques such as sound zone control (SZC), spatial active noise control (ANC), and immersive audio reproduction are increasingly deployed in practical systems including personal audio devices, cars, and smart environments. These methods shape acoustic sound fields using multiple loudspeakers to enhance desired audio in specific regions while suppressing it elsewhere.

In practice, robust spatial audio control remains challenging because most methods rely on control filters computed offline from pre-measured acoustic impulse responses (IRs) that are assumed fixed during deployment. However, environmental changes such as listener movement, transducer drift, or temperature variations alter the IRs and degrade performance \cite{betlehem_temperature_2018,moller_influence_2019,coleman_acoustic_2014,bhattacharjee_robust_2025}. Among these factors, variations in the speed of sound are particularly critical, as they introduce systematic delay and phase mismatches that severely affect spatial control \cite{bhattacharjee_sound_2025,betlehem_temperature_2018,caviedes_nozal_effect_2019,
}. Nevertheless, only a few approaches address sound speed variations \cite{bhattacharjee_sound_2025,betlehem_temperature_2018,heuchel_large-scale_2020, brunnstrom_spatial_2025, zhang_cgmm-based_2023}.
Constraint-based IR reshaping \cite{betlehem_temperature_2018}, learned parametric propagation models \cite{heuchel_large-scale_2020}, learned IR priors \cite{zhang_cgmm-based_2023}, and covariance-prior-based methods \cite{brunnstrom_spatial_2025} have been proposed to improve robustness. However, these approaches typically require repeated calibration, multiple microphones during deployment, or substantial pre-measured training data. Adaptive filtering and secondary path modeling can track acoustic changes online \cite{hu_sound_2023,zhang_alternating_2025}, but require access to microphone signals at all control points, limiting applicability in minimally instrumented systems. 
Recently, \cite{bhattacharjee_sound_2025} proposed interpolation-based IR modeling using a Sinc Interpolation–Compression/Expansion Resampling (SICER) framework, which enables recomputation of control filters at new sound speeds. However, this assumes that the sound speed is known or estimated separately, for example using temperature and humidity sensors. 
More broadly, classical sound speed estimation methods are often coupled with source localization and rely on multiple spatially distributed microphones \cite{annibale_tdoa-based_2013,mahajan_3d_2001,hu_simultaneous_2011,othmani_review_2024}, or require dedicated measurement procedures \cite{godin_passive_2014,anderson_direct_1998}, which is poorly suited for online adaptation in systems with limited sensing infrastructure.

In this paper, we propose an online sound speed estimation method that operates during multichannel audio playback using only a single observation microphone, not necessarily placed at the control points. The method exploits the structured effect of sound speed variations on acoustic propagation, as described in \cite{bhattacharjee_sound_2025}, and estimates the sound speed directly from the reproduced audio signal without additional sensors. As an example application, the estimated sound speed is integrated into an SZC framework to compensate for propagation mismatches without modifying the underlying control architecture. The proposed approach enables practical, single-channel sound speed tracking for robust spatial audio control.

\section{Frame-based Signal Model}

We consider a general frame-based multichannel audio system where a set of loudspeakers, $\mathcal{L}$, is used to generate a desired sound field at a set of control-point microphones $\mathcal{M}$. For example, in SZC the loudspeakers are used to create different sound zones with different desired sound fields, cf.~Section~\ref{sec:szc_background}~\cite{galvez2015time, lee2018unified}. 
We denote by ${\bm{h}_{m,l} \in \mathbb{R}^{K }}$ the IR from the $l^{\text{th}}$ loudspeaker to the $m^{\text{th}}$ microphone. Each loudspeaker is assumed to be equipped with a finite impulse response (FIR) filter, $\bm{q}_l\in \mathbb{R}^J$ to control the reproduced sound field, e.g., an ANC or SZC filter. Then, for an input signal frame $\bm{x}[\tau]=[x[\tau N - N +1], \ldots, x[\tau N]]\in \mathbb{R}^N$, the signal frame reproduced by all $L$ loudspeakers at microphone $m$ and frame index $\tau$, is
\begin{equation}
        \bm{p}_m[\tau]
        =\sum_{l\in \mathcal{L}}h_{m,l}\ast y_l[\tau]
        =\sum_{l\in \mathcal{L}}h_{m,l} \ast q_l\ast x[\tau] \in \mathbb{R}^N,
\end{equation}
where $\ast$ denotes convolution, 
and $\bm{y}_l[\tau]\in \mathbb{R}^N$ is the $l^{\text{th}}$ loudspeaker output signal for frame $\tau$. 
We use overlap-add and buffers for each convolution to avoid frame boundary errors\cite{christensen_introduction_2019}, and let all signal frames have length $N$. 
We assume ${K-1 \leq N}$ and $J-1 \leq N$ such that the convolution tail can be stored in a single frame buffer\cite{bhattacharjee_robust_2025}.
For a vector $\bm{v} \in \mathbb{R}^J$ containing time-consecutive samples of a signal, $v[j]$, we define a buffering operator, $\operatorname{Buff}^N_{K-1}(\bm{v})$, which extracts the last $K-1$ elements of $\bm{v}$ and zero-pads to length $N$ as
\begin{equation}
    \operatorname{Buff}^N_{K-1}(\bm{v}) \triangleq [(\bm{v})_{J-(K-2):J}^\mathrm{T}, \underbrace{0,\ldots,0}_{N-(K-1)}] \in \mathbb{R}^{N},
\end{equation}
where $(\bm{v})_{a:b} \triangleq [v[a], v[a+1], \ldots, v[b]]^\mathrm{T}$. 
The reproduced signal from loudspeaker $l$ at microphone $m$ in frame $\tau$ is then
\begin{equation}
      \bm{p}_{m,l}[\tau]= (h_{m,l} \ast y_l[\tau])_{0:N-1} + \operatorname{Buff}^N_{K-1}(h_{m,l} \ast y_l[\tau-1]).
\end{equation}
Similarly, we can express $\bm{y}_l[\tau]$ from $\bm{x}[\tau]$ and $\bm{q}_l$.


\section{Online Sound Speed Estimation}
We first present our model for sound speed changes in the reproduced audio based on the SICER model for speed changes on IRs~\cite{bhattacharjee_sound_2025,
}.
\subsection{Modeling Sound Speed Effect on IRs \& Signals}
Following the SICER model in \cite{bhattacharjee_sound_2025}, an IR measured at sound speed $\cold$, $\bm{h} \in \mathbb{R}^K$, can be mapped to a new sound speed $\cnew$ with IR, $\hat{\bm{h}} \in \mathbb{R}^{I}$
\begin{equation}
    \hat{\bm{h}}^{(\alpha)} = \alpha \bm{S}^{(\alpha)}\bm{h}, \quad \alpha=\frac{\cnew}{\cold}
\end{equation}
where 
$\bm{S}^{(\alpha)} \in \mathbb{R}^{I \times K}$ is a sinc interpolation matrix with the $i^{\text{th}}$ row defined as
${\bm{s}_i^{(\alpha)} = \left[\sinc\left(\alpha i \right), \ldots, \sinc\left(\alpha i - K+1\right)\right]}$.

Using this model and assuming a uniform temperature in the environment, the reproduced signal under a sound speed change becomes
\begin{equation}
\begin{aligned}
        \hat{\bm{p}}^{(\alpha)}_m[\tau]
        =\sum_{l\in \mathcal{L}} &\left(\hat{h}_{m,l}^{(\alpha)} \ast y_l[\tau]\right)_{0:N-1} \\&+ \operatorname{Buff}^N_{K-1}\left(\hat{h}_{m,l}^{(\alpha)} \ast y_l[\tau-1]\right)\label{eq:reproduced_sicer_model}
\end{aligned}
\end{equation}
With this model for audio reproduction under new sound speeds, we can now estimate the change in sound speed based on recordings of reproduced audio.

\subsection{Estimating Sound Speed}
We now present our method for estimating the current sound speed, $\cnew$, online during audio playback. 
We let $m_o$ be an observation microphone in the environment, where we assume the IR, $\bm{h}_{m_0,l}$, from each loudspeaker to the observation microphone is known at a reference sound speed, $\cold$, e.g., measured prior to audio playback. 
Letting $\bm{p}_m^{\text{meas}}[\tau]$ be the signal measured at microphone $m$ for frame $\tau$, we can estimate the current sound speed by minimizing the error between the measured signal and the modeled reproduced signal under a sound speed change in \eqref{eq:reproduced_sicer_model}, i.e.,
\begin{equation}
    \hat{c}_{\text{new}} = \arg \min_{c_{\text{new}}}
    \left\| \bm{p}_{m_o}^{\text{meas}}[\tau] - \hat{\bm{p}}_{m_o}^{(\alpha)}[\tau] \right\|_2^2. \label{eq:mse_cost}
\end{equation}
The resulting single-parameter, non-convex, nonlinear optimization problem can be readily solved numerically. This estimation is then performed frame-wise during playback. We note that the cost function can be generalized to cover multiple microphones; however, as we will see in Section~\ref{sec:experiments}, using a single microphone is already effective.



\section{Robust Control with Sound Speed Changes}

In this Section, we show our complete algorithm for sound speed estimation along with robust spatial audio control. As an example of a spatial audio control system, where filters depend on the IRs and thereby sound speed, we consider SZC.

\subsection{Sound Zone Control Background}\label{sec:szc_background}

We consider a SZC system as depicted in Fig.~\ref{fig:szc_setup}, where the goal is to find a set of loudspeaker control filters $\{\bm{q}_l\}_{l\in \mathcal{L}}$, that creates a bright zone (BZ) with a desired sound field and a dark zone (DZ) with silence. When computing the filters prior to deployment, each of the control zones are equipped with a set of control microphones, i.e., $\mathcal{M}_B$ and $\mathcal{M}_D$. For each of these control points, we assume the IRs from each of the loudspeakers are known at a reference sound speed, $\cold$, and compute filters based on these. Consequently, the filters are only optimal for that specific sound speed. 
Furthermore, for simplicity and as is common in SZC filter design, we assume the input signal is $x[n]=\delta[n]$, i.e., a deterministic spectrally white signal \cite{galvez2015time,lee2018unified}. The sound reproduced at the $m^{\text{th}}$ microphone in either the BZ or DZ microphones can then be expressed as the vector
${\bm{p}_m= \sum\nolimits_{l=1}^{L} {h}_{m,l} * {q}_l=\sum\nolimits_{l=1}^{L} \bm{H}_{m,l}\bm{q}_l\in \mathbb{R}^{K+J-1}}$, where $\bm{H}^{(s)}_{m,l}$ is the convolution matrix containing the IR $\bm{h}_{m,l}$\cite{galvez2015time}.
The reproduced sound field at all microphones in, e.g., the BZ is then
\begin{flalign}
  \bm{p}_B = \left[\bm{p}_1, \ldots, \bm{p}_{M_B}\right]^{\textrm{T}} &= \bm{H}_B \bm{q} \in \mathbb{R}^{M_B(K+J-1)}.
\end{flalign}
Here ${\bm{H}_B=\left\{\bm{H}_{m,l}\right\}_{m \in \mathcal{M}_B, l \in \mathcal{L}}}$ 
and 
$ {\bm{q}= \left[\bm{q}_1^{\textrm{T}}, \ldots, \bm{q}_{L}^{\textrm{T}}\right]^{\textrm{T}}} \in \mathbb{R}^{LJ}$\cite{galvez2015time}. Letting $\bm{d}_m\in \mathbb{R}^{K+J-1}$ be the desired signal at microphone $m$, defined according to a virtual source position, $z$,~\cite{galvez2015time}, we stack these into a desired signal vector for the BZ as $\bm{d}_B = \left[\bm{d}_1^\mathrm{T}, \ldots, \bm{d}_{M_B}^\mathrm{T}\right]^\mathrm{T} \in \mathbb{R}^{M_B(K+J-1)}$. The desired signal for the DZ is silence, i.e., $\bm{d}_D = \bm{0}\in \mathbb{R}^{M_D(K+J-1)}$.
\subsubsection{VAST Approach} The filters are derived by minimizing the weighted mean-squared error between the desired and reproduced signals 
\begin{align}
    \xi(\bm{q})
    = \bm{q}^\mathrm{T} \bm{R}_B\bm{q} + \mu \bm{q}^\mathrm{T}\bm{R}_D\bm{q} - 2 \bm{q}^\mathrm{T} \bm{r}_B + \left\lVert\bm{d}_B\right\rVert_2^2,
\end{align}
where $\mu$ is a weighting parameter, $\bm{r}_B = \bm{H}_B^\mathrm{T} \bm{d}_B$, and ${\bm{R}_B = \bm{H}_B\bm{H}_B^{\textrm{T}}}$ and $\bm{R}_D = \bm{H}_D\bm{H}_D^{\textrm{T}}$ are the spatial covariance matrices corresponding to the BZ and DZ\cite{lee2018unified}. Using the Variable-Span Trade-off (VAST) approach\cite{lee2018unified}, the solution is found via the eigenvalue decomposition of $\bm{R}_D^{-1}\bm{R}_B$ and yields a $V$-rank ($1 \leq V \leq LJ$) approximation of $\bm{q}$ as
\begin{equation}
    \bm{q} = \sum_{v=1}^{V} \frac{\bm{u}_v^\mathrm{T} \bm{r}_B}{\lambda_v + \mu} \bm{u}_v,
\end{equation}
where $\lambda_1  \geq \ldots \geq \lambda_V$ are the non-negative real-valued eigenvalues of $\bm{R}_D^{-1}\bm{R}_B$, $\mu$ controls the weighting between BZ and DZ performance, and $V$ trades-off between signal distortion and acoustic contrast\cite{lee2018unified}.
For positive semi-definite $\bm{R}_D$, regularization can be applied as ${\bm{R}'_D=\bm{R}_D + \gamma \bm{I}}$.
We can now combine the SZC filter computation with our online sound speed estimation to a robust control algorithm.

\subsection{Sound Speed Estimation and Robust Control Algorithm}
Algorithm~\ref{alg:speed_estimation} summarizes the proposed procedure with two main components: $1)$ online sound speed estimation and $2)$ updating the control filters when the speed change exceeds a given threshold $c_{\text{thresh}}$ based on SICER adjusted IRs \cite{bhattacharjee_sound_2025}.
Sound speed is estimated frame-wise using either a full grid search or an adaptive grid search around the previous estimate.


\algblock[Name]{Initialization}{}
\algsetblock[Name]{Initialization}{}
{}{0cm}
\begin{algorithm}[t]
\caption{Online Sound Speed Estimation using SICER}
\label{alg:speed_estimation}
\begin{algorithmic}[1]
\State \textbf{Input:} $\cold$, $\{\bm{h}_{m,l}\}_{m \in m_o \cup \mathcal{M}_B \cup \mathcal{M}_D, l\in \mathcal{L}}$, $c_{\text{thresh}}$
\State search range $[c_{\min}, c_{\max}]$, step size $\Delta c$,
\State \textbf{Optional Input:} Adaptive search range $c_{\text{width}}$, $\Delta c_{\text{adapt}}$
\State \textbf{Initialization} Compute $\{\bm{q}_l[1]\}$ for $\cold$, $\{\bm{y}_l[0]=\bm{0}\}$, $c_{\text{filt}} = \cold$,

\For{$\tau=1,2, \ldots $}
    \State Compute \& play $\{\bm{y}_l[\tau]\}$
    \State Measure $\{\bm{p}_m^{\text{meas}}[\tau]\}_{m \in \mathcal{M}_O}$
    %
    \If{Adaptive search range \text{\&} $\tau > 1$}
        \State $c_{\min} = \max\{ \hat{c}_{\text{new}}[\tau-1] - c_{\text{width}}, c_{\min}\}$
        \State $c_{\max} = \min\{\hat{c}_{\text{new}}[\tau-1] + c_{\text{width}}, c_{\max}\}$
        \State $\Delta c =\Delta c_{\text{adapt}}$
    \EndIf

            
    \State $ \hat{c}_{\text{new}}[\tau] = \arg \min\limits_{\cnew}
    \left\lVert \bm{p}_m^{\text{meas}}[\tau] - \hat{\bm{p}}_m^{(\alpha)}[\tau] \right\rVert_2^2$
    \Statex $\qquad\ \text{s.t.}~\alpha = \frac{\cnew}{\cold},~ \cnew \in  [c_{\min}, c_{\max}] ~\text{with step size}~\Delta c$\vskip 3pt
    \If{$\lvert \hat{c}_{\text{new}}[\tau] - c_{\text{filt}} \rvert \geq c_{\text{thresh}}$
    } \Comment{Update filters}  \vskip 3pt
     \State $\hat{\alpha} = \frac{\hat{c}_{\text{new}}[\tau]}{\cold}$
        \State $\hat{\bm{h}}_{m,l}^{(\hat{\alpha})}=\hat{\alpha} \bm{S}^{(\hat{\alpha})}\bm{h}_{m,l}, \quad m\in \mathcal{M}_B \cup \mathcal{M}_D,~l\in \mathcal{L}$
        \State Compute new filters $\{\bm{q}_l[\tau+1]\}$ from $\{\hat{\bm{h}}_{m,l}^{(\hat{\alpha})}\}$
        \State $c_{\text{filt}} = \hat{c}_{\text{new}}[\tau]$
    \EndIf
        
\EndFor

\end{algorithmic}
\end{algorithm}




\section{Experimental Evaluation}\label{sec:experiments}
\subsection{Simulation Setup}
\begin{figure}[t]
    \centering
    \includegraphics[width=.6\linewidth, clip, trim={0 0 0 2mm}]{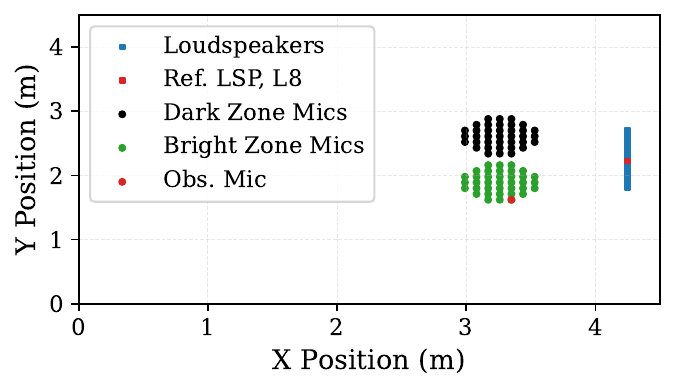}
    \vspace*{-3mm} 
    \caption{Simulation setup for evaluating online sound speed estimation and robust sound zone control performance. }
    \vspace{-3mm}
    \label{fig:szc_setup}
\end{figure}

To evaluate both tracking and robust SZC performance, we simulate a setup similar to~\cite{lee_fast_2021,bhattacharjee_sound_2025} with the RIR Generator Toolbox\cite{habets_rir_2014
} and sample rate of $16$\,kHz. The setup has a square room of size $4.5 \times 4.5 \times 2.2$\,m, cf.~Fig.~\ref{fig:szc_setup}, with a linear array of $L=16$ loudspeakers spaced $6$\,cm apart, and a BZ and DZ each with $M_B=M_D=37$ control microphones and $9$\,cm spacing. All transducers are placed at a height of $1.2$\,m. To facilitate faster processing, we simulate a short reverberation time of $RT_{60}=100$\,ms, giving an IR length of $K=800$. IRs are simulated with sound speeds ranging from $333$\,m/s to $353$\,m/s with $2$\,m/s steps, and the IRs at these steps are considered as the true ground truth (GT). The SZC filters are computed with length $J=500$, weighting $\mu=1$, no regularization, and with loudspeaker $l=8$ as reference desired source for the BZ. As observation microphone we use a point on the edge of the BZ.

We consider $22$\,s input signals with varying spectral content: white noise, speech (four talkers from EARS\cite{richter2024ears}) and instrumental rock music (MUSAN\cite{musan2015}). 
The frame length is $N=4000$ ($250$\,ms) with no overlap.

To stress-test the method, sound speed is increased by $2$\,m/s (corresponding to approximately $1.7\,^\circ\text{C}$) every $8$ frames by changing the GT propagation IR\footnote{Results from other speed changes can be found on our GitHub along with our code: \url{https://github.com/afuglsAAU/EUSIPCO2026SoundSpeedEst}}.
We use a fixed search range of $[326,360]$\,m/s with step size $0.25$\,m/s, and adaptive search range width $\pm 3$\,m/s with step size $0.1$\,m/s. The filter update threshold is set to $1$\,m/s.



\subsection{Tracking Performance}
\begin{figure*}[t]
    \centering
    \includegraphics[width=\linewidth]{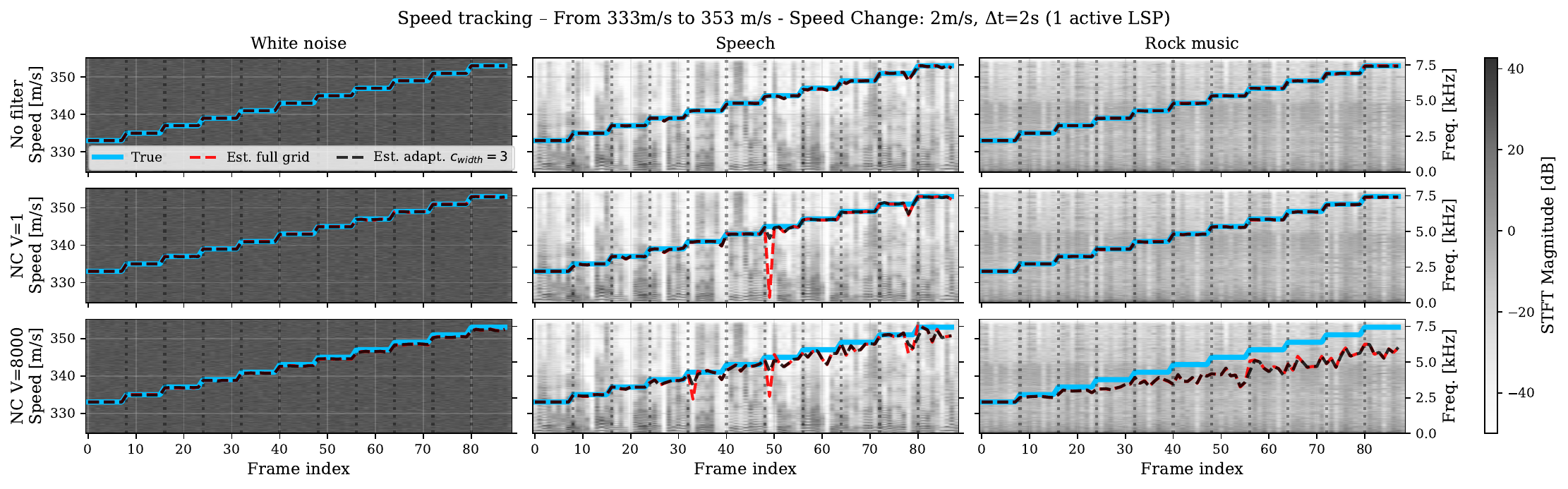}
    \caption{Tracking performance with one active loudspeaker overlaid with spectrogram of non-filtered input signal. Showing the proposed (red: full grid \& black: adaptive grid) and the true sound speed (blue) for different audio signals (columns), both without filtering and with non-speed-corrected (NC) filters for two different ranks (rows). }
    \label{fig:track_one_lsp_up}
\end{figure*}

To focus only on tracking performance, we first consider the simple case where only the reference loudspeaker is active, both when no filter is applied, i.e., ($\bm{q}_8=[1,0,\ldots,0]$), and when applying a SZC filter but with no speed correction (NC), i.e., keeping it fixed across time. The results in Fig.~\ref{fig:track_one_lsp_up} show accurate tracking for white noise and speech, and robust performance for music despite its more limited spectral diversity. Tracking accuracy decreases when the input signal has low energy or reduced frequency content, e.g., around frames $32$, $50$ and $78$ for speech input. The adaptive search range improves stability in such frames. Furthermore, applying full rank ($V=8000$) pre-filtering decreases performance slightly for speech and white noise but more severely for rock music. Overall, the proposed method reliably tracks sound speed using only a single observation microphone.

\begin{figure*}[t]
    \centering
    \includegraphics[width=\linewidth]{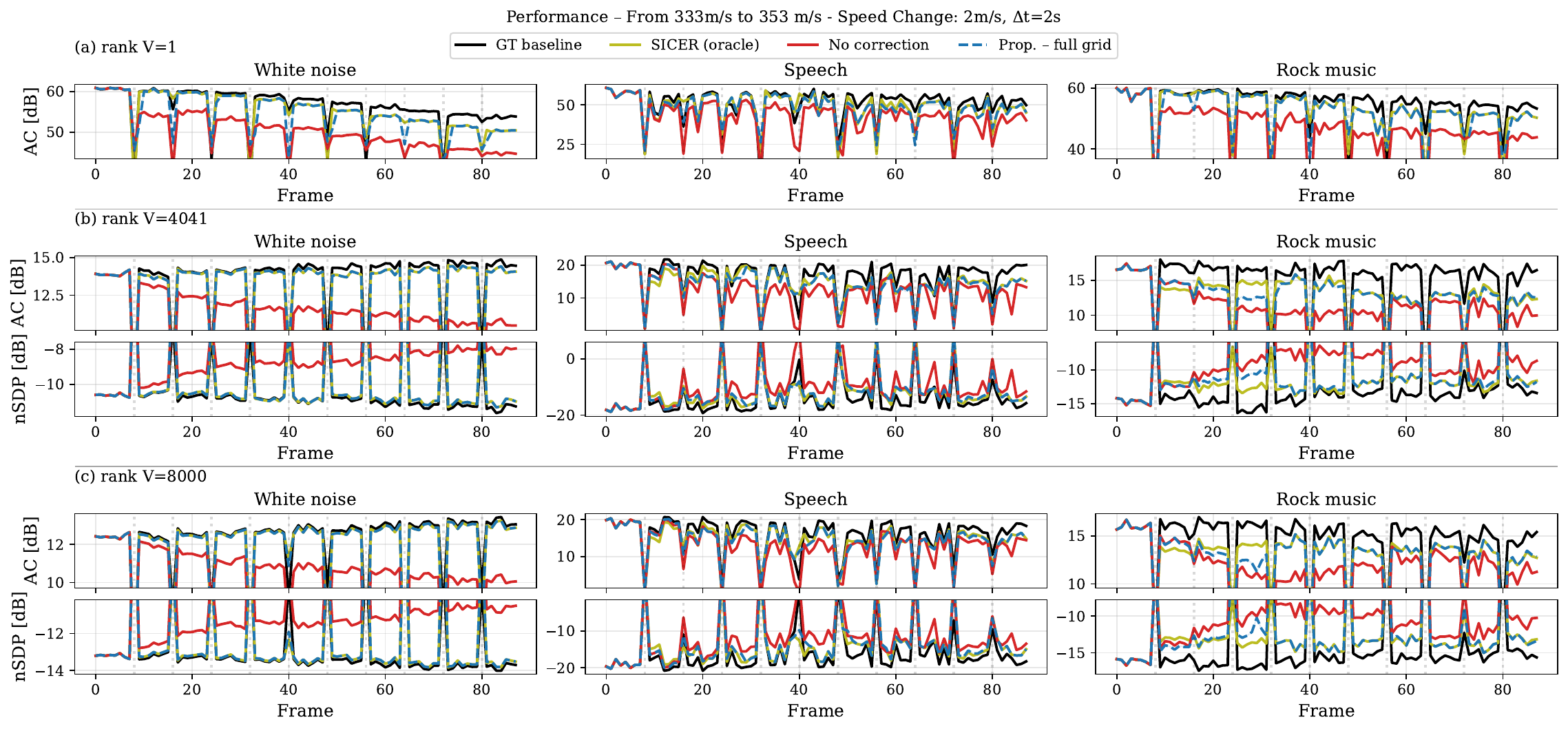}
    \caption{SZC for VAST ranks $V=1$ (a), $V=4041$ (b) and $V=LJ=8000$ (c) and different input signals (columns). For visibility the vertical axes are scaled according to performance between frames with speed changes. The proposed (blue dashed) is compared to uncorrected filters (red), oracle SICER (green) and GT performance (black).}
    \label{fig:performance}
\end{figure*}





\subsection{Sound Zone Control Performance}
We next evaluate the SZC performance of the proposed algorithm when all loudspeakers are active 
in terms of the acoustic separation between the zones and the reproduction error in the BZ. These are quantified, respectively, by the acoustic contrast (AC) and normalized signal distortion power (nSDP), defined as
\begin{align}
    \textrm{AC}[\tau] &= 10\log_{10}\left(\frac{M_D}{M_B}\frac{\lVert \bm{p}_B[\tau]\rVert^2}{\lVert \bm{p}_D[\tau]\rVert^2}\right),\\
    \textrm{nSDP}[\tau] &= 10\log_{10}\left(\frac{\lVert \bm{d}_B[\tau]-\bm{p}_B[\tau]\rVert^2}{ \lVert \bm{d}_B[\tau]\rVert^2}\right). \label{eq:nSDP}
\end{align}
We compare performance against fixed filters (no speed correction) (lower baseline), filters computed with ground truth IRs (upper baseline) and oracle SICER\cite{bhattacharjee_sound_2025}. 

Figure~\ref{fig:performance} shows the SZC performance for the different input signals and three VAST ranks. nSDP for rank $V=1$ is omitted for brevity as it is close to $0$\,dB for all inputs and methods. We note that the sharp changes are caused by the simulated step change in speed but in practice the change would be smooth. The results show the proposed method approaches the GT baseline across signals and VAST ranks and closely matches oracle SICER, showing the remaining gap to the GT is due to SICER interpolation, not estimation error.
Hence, sound speed can be accurately tracked during multi-channel SZC, while improving SZC performance against non-correcting fixed filters.


\section{Conclusion}

We presented an online method for estimating sound speed by minimizing the mismatch between the reproduced audio signal and a sinc interpolation–compression/expansion resampling-based acoustic model. 
The approach estimates sound speed directly from the ongoing playback using only a single observation microphone, 
without additional sensors or dedicated calibration measurements.
Simulation results demonstrated accurate tracking across different input signals and in the presence of spatial audio control filters. Integrating the estimated sound speed into a sound zone control framework enabled compensation for propagation mismatches and improved performance compared to fixed-filter designs, even under exaggerated sound speed variations.
Future work includes improving computational efficiency and validating the method using measured impulse responses.
\section{REFERENCES}
\printbibliography[heading=none]

\end{document}